\documentstyle{article}

\def\Hz{{\rm Hz}}
\def\Gs{{\rm Gauss}}
\def\Jan{{\rm Jansky}}
\def\chg{\leftrightarrow}
\def\Lc{L_{\rm coh}}
\def\GW{\small\rm GW}
\def\EW{\small\rm EW}
\def\ei{e^{i[\frac 12 (k_+ +k_-)x - \omega t]} }
\def\eim{e^{i(k_- x - \omega t)} }
\def\eip{e^{i(k_+ x - \omega t)} }

\begin{document}
\title{PROMPT AND DELAYED RADIO BANGS AT KILOHERTZ BY SN 1987A:\\[5pt]
       A TEST FOR GRAVITON-PHOTON CONVERSION}
\author{Daniele Fargion \\
       {\small Physics Dept., Universit\`a degli Studi di Roma
        ``La Sapienza''and INFN,} \\ 
        {\small P. le A. Moro 2, 00185 Rome, Italy.}}
\date{8 August 1995}
\maketitle
\begin{abstract}
A sequence of prompt and delayed radio signals at tens of kilohertz
should reach the Earth (or Jupiter) due to graviton--photon conversion
in interstellar as well as local planetary magnetic fields. These radio
fluxes may be a detectable probe of a huge gravitational burst expected
from Supernovae explosions.
The earliest prompt radio signal, coinciding with the neutrino burst, is
due to conversion in the terrestrial (or Jovian) magnetic field and is
below the micro-Jansky (or milli-Jansky) level for a galactic Supernova
like SN1987A. A later radio signal, a ``tail'', due to the same graviton -
radio wave conversion in random interstellar fields will maintain a relic
radio ``noise'' for hundreds or thousands of years and might even be still
detectable by a very sensitive network of satellite antennas at the
kilohertz band. Exact solutions are presented here for the graviton-photon
conversion in a refractive medium, as well as their consequences for high
energy supernovae and the 2.726 K background radiation.
\end{abstract}
% ] %%%%%%%%%%%%%%%%%%%%%%%%%%%%%%%%%%%%%%
% \email{1}{fargion@roma1.infn.it}

\section{Introduction: the graviton--photon conversion}

When a massless graviton interacts with an orthogonal component of a
stationary electromagnetic field, it may decay into a pair of massless
photons: a real photon with almost the same graviton wavevector and
frequency, and a virtual one due to the external stationary field. Since
Gravitational Waves (GW) and Electromagnetic Waves (EW) propagate in
perfect vacuum at the speed of light $c$, the processes reinforce
themselves, leading to resonant phenomena discovered first by Gertsenshtein
in 1961 [1]. There is an analogous and better known process in general
relativity, with a Feymann diagram describing photon-photon scattering
by a virtual graviton: the deflection of light by gravitational field which
made so popular, since 1919, the Einstein theory.  However,
contrary to light deflection, graviton-photon conversion has
been never observed and may have deeper consequences in future
GW astronomy. In classical field theory one may understand
graviton-photon conversion as a result of a background metric perturbation
in a stationary magnetic or electric field: the space-time vibration of
GWs squeezes the magnetic or electric field lines, which themselves become
more and more, all along the GW propagation, sources of real EWs
[2,3].  The GW-EW conversion is a reversible process.  Indeed, a
real photon in the presence of an external magnetic (or electric) field may
be annihilated by a virtual photon, creating a real graviton. This reverse
process could be seen from classical field theory as a constructive
interference between free EW and the stationary electromagneticel field: this
interference is the source of an ``energy beating component'' along the
EW propagation path, which reinforces itself and becomes a
``resonant'' source of a GW flying in the same direction. At a
very long distance the full process is oscillatory, but for most
realistic astrophysical and even cosmological cases the conversion occurs
only at a very limited size; then it may be important to consider only
partial conversion from one (GW) to the other (EW) form of energy,
with the efficiency depending quadratically on the distance. This reversible
dimensionless energy conversion efficiency $\alpha$ is equal in
both directions (GW $\chg$ EW), and in perfect vacuum, in the first
approximation, was found [1,6] to be
\begin{equation}
\alpha_a = \frac{P_{\rm GW}}{P_{\rm EW}}
   = \frac{GB^2L^2}{c^4}= 8.26 \cdot 10^{-25}
\left(
\frac{B}{10^6 G}\ \frac{L}{\rm 30 km} \right)^2.    %\eqno(1)
\end{equation}
Here $P_{\GW}$ and  $P_{\EW}$ stand for the GW and EW powers.

In what follows we shall consider for real cases only stationary
magnetic fields because, as Nature and consequentely the Maxwell
equations teach us (and Parker's bounds on monopoles imply), we may expect
only large-size coherent magnetic fields due to the total (or negligible)
absence of free magnetic monopoles.

Therefore $B \equiv |B_{\perp}|$ is a
stationary magnetic field orthogonal to the free GW (or EW) and $L$ is the
path length crossed by GW (or EW) under the stationary field $B$;
$G$ and $c$ are the Newtonian constant and the velocity of light,
respectively.

When the GW-EW conversion takes place in an oscillatory manner (in
perfect vacuum), the conversion efficiency becomes $\alpha_0$:
\begin{equation}
\alpha_0 = \frac{P_{\rm GW}}{P_{\rm EW}}=
\tan^2 \left[\frac{G^{1/2}BL}{c^2}\right]
\simeq \left( \frac{G^{1/2}BL}{c^2}\right)^2 = \alpha_a.   %\eqno(2)
\end{equation}
The last approximate relation holds only when $\alpha_0 \ll 1$, as was
assumed in equation (1).
The smallness of the above conversion efficiency makes the phenomena
very difficult to observe in laboratory. Moreover, the presence of any
realistic refractive index (in a terrestrial laboratory or in an
astrophysical or even cosmological framework) often reduces the efficency,
or, worse, it may dilute its arrival in time, leading to a serious problem
concerning the signal-to-noise ratio.  Therefore it is more convenient to
disregard the very difficult double conversion (photon-graviton-photon) in
laboratory and to consider only the GW-EW conversion in an astrophysical
framework.

The presence of a refractive index will lead to a sequence of EW
$\chg$ GWs conversions, an incoherent ``multiconversion" which
enhances the oscillatory conversion.

One could treat as a GW source one of the few known periodic sources, such
as the binary system PSR\,1913+16. However, their frequency is very low,
so that the refractive index of free electrons in the interstellar
space makes the GW-EW conversion hopelessly weak.

Therefore the best sources we are able to suggest for the GW-EW conversion
are the galactic Supernovae which might emit at higher total power and at
much larger frequency, so that the refractive index of electrons might be
less severe in screening or delaying the GW $\chg$ EW conversion.

Moreover, the local terrestrial or Jovian magnetic fields might also convert
SN GWs as soon as neutrino bursts arrive, and therefore it would be easier
to observe them in coincidence with time-direction constraints. Extended
interstellar magnetic fields may give life to much longer delayed radio
tails. This occurs in the presence of free intergalactic charges, because
the effective ``mass of radio photons'' makes the photons travel at a speed
smaller than $c$. Therefore the radio signal will come from long distances
later and later, leading to a diluted radio signal. Their continuous
conversion signals will produce a long (thousands of years) delayed radio
noise which might be more difficult to observe.

The GW conversion cannot take place too near a supernova where the
magnetic fields are the greatest, because of the huge ionization (and
hence a huge refractive index) due to the explosion. Unfortunately,
the stellar and sky radio noise at tens of kilohertz may be dominated by
local sources of noise which can hide the simplest GW $\chg$ EW conversion
in the perfect case.

Finally, the EW $\chg$ GW conversion may also affect (but at
much lower level than the present sensitivity) the cosmic background
rediation, at a temperature perturbation level $\Delta T/T \simeq
10^{-7}$.

\section{Gravitational and electromagnetic field equations in stationary
fields}                        % S2
The photon-graviton conversion phenomenon has been first discovered
by Gertsenhtein [1] in 1961. This process is a secondary effect of
gravitational synchrotron radiation [2,6] and has been
also analyzed by different authors [3,6]. Let us
briefly reconsider the equations of the GW $\chg$ EW oscillations.
Following Landau and Lifshits's notation [7], we consider a
nearly flat space-time $\eta_{ik}$ with a small metric perturbation
$h_{ik}$; we also introduce the traceless tensor $\psi^i_k$:
\begin{equation}
g_{\mu\nu}=\eta_{\mu \nu}+ h_{\mu \nu};  \;\;\;\;\
h_{\mu\nu}=\psi_{\mu \nu}-\frac{1}{2}\psi^{\sigma}_{\sigma}g^0_{\mu\nu}
,%\eqno(3)
\end{equation}
so that the Einstein field equations $ G_{\mu \nu}=(8 \pi G/c^4)T_{\mu \nu}$
in the linear approximation become
\begin{equation}
\Box \psi^{\mu}_{\nu}= \Box h^{\mu}_{\nu}
= - \frac{16 \pi G}{c^4} \tau^{\mu}_{\nu}        %\eqno(4)
\end{equation}
where $\tau^{\mu}_{\nu}$ is the energy momentum tensor (the first equality
in (4) holds because $\psi^{\mu}_{\mu}=0$). In an external stationary
electromagnetic field $F^{\sigma \tau(o)}$ and in the presence of a free EW,
$\tilde{F}^{\sigma \tau}$, the total electromagnetic (EM) field is
\begin{equation}
F^{\mu \nu}\equiv F^{\mu \nu(0)}+ \tilde{F}^{\mu \nu}   %\eqno(5)
\end{equation}
The corresponding energy-momentum tensor is
\begin{equation}
\tau^{\mu}_{\nu}=\frac{1}{4\pi} \left[F^{\mu \sigma} F_{\nu \sigma}-\frac{1}{4}
   \delta^{\mu}_{\nu}\left(F^{\sigma\tau}F_{\sigma\tau}\right)\right] 
  %\eqno(6)
\end{equation}
This expression
may be considered to consist of three independent components: the first
one proportional to constant terms, $\left(F^{\mu \sigma(0)} F_{\mu
\nu (0)}\right)$, the second one proportional to the free field terms, $\left(
\tilde{F}^{\mu \nu} \tilde{F}_{\mu \tau}\right)$ describing the massless EW (or
just photons) and the third one proportional to the free-stationary field
interference: $2\tilde{F}^{\mu \nu} F^{(o)}_{\mu \sigma}$.

The first term is not a GW source since it is constant, the second one
cannot be a source because of the energy-momentum conservation law of free
massless particles, while the third one is resonant and can be an effective
source of GWs. Indeed, the presence of an external (virtual
photon) field allows the necessary momentum to be transferred outside the
free massless system.
Then we can write the effective Einstein equations as follows:
\begin{eqnarray}        
\Box h^{\mu}_{\nu} & = & - \frac{16 \pi G}{c^4} \tau^{\mu}_{\nu}\;=\nonumber \\
                   & = & - \frac{8G}{c^4}
\left[F^{(0)\mu \sigma} \tilde{F}_{\nu \sigma}- \frac{1}{4} \delta^{\mu}_{\nu}
\left(F^{(0)\sigma \tau} \tilde{F}_{\sigma \tau} \right)\right]  %\eqno(7)
\end{eqnarray}
Due to the absence of magnetic monopoles we shall consider mainly large-size
magnetic fields. For simplicity let us consider a uniform
stationary magnetic field $B_{0z}$ along the $z$ axis and an orthogonal EW,
polarized with an induction vector $\tilde{B}_z$, parallel to $B_{0z}$ and
propagating along the $x$ axis. In this case the tensor components of the
wave equations (8) reduce to
\begin{eqnarray}
\Box h^1_1 = \Box h^2_2=\Box h^1_0= (4G/c^4) B_{0z} \tilde{B_z}, \nonumber \\
\Box h^3_3 = \Box h^0_0=\Box h^0_1 =-(4G/c^4) B_{0z} \tilde{B_z}; %\eqno(8)
\end{eqnarray}
for the other metric components one finds
\begin{equation}
h^2_2=-h^3_3,   \;\;\;\;   h^3_2=h^2_3=h^3_2=0.         %\eqno(9)
\end{equation}
Moreover, for simplicity, we may assume the EW to be a plane linearly
polarized wave
\begin{equation}
\tilde{B_z} \equiv \tilde{B}_{z0} \; e^{i(kx-\omega t)} \;\;\;\;\;\;
(|\tilde{E_y}|=|\tilde{B_z}|).                                 %\eqno(10)
\end{equation}
Neglecting the feedback reaction (of GW on EW), one can solve
Eq.\,(8) [1] assuming a slowly varying function
$h^2_2=b(x)$, $\left(d^2 b/dx^2=0\right)$. This procedure leads to a
conversion factor $\alpha_a$ of Eq.\,(1); in general one must also include
the reverse process, i.e., EW production by GW. Let us evaluate this
general behaviour. The Maxwell equations in a vacuum curved space-time
\begin{equation}
F_{\mu \nu; \sigma}+ F_{\sigma \mu; \nu}+F_{\nu \sigma;\mu}=0  %\eqno(11)
\end{equation}
may be reduced in the present case to a few relevant components by
substituting the covariant derivative definition $(F_{\mu \nu; \sigma}=
F_{\mu \nu,\sigma}-\Gamma^{\tau}_{\mu \sigma} F_{\tau \nu}-\Gamma^{\tau}_{\nu
\sigma} F_{\nu \tau})$ [3,4]:
\begin{equation}
F^{21}_{,0}+F^{20}_{,1}=-B_{oz}h_{22,1},         \;\;\;\;
F^{21}_{,1}+F^{20}_{,0}=0.                       %\eqno(12)
\end{equation}
Taking a space derivative in the first equation and a time derivative in the
second one, we reduce Eqs.\,(8) and (12) to a set of wave equations:
\begin{eqnarray}
\Box \tilde{B}_z=-B_{0z}h_{22,11}=B_{0z}k^2 h_{22}, \nonumber   \\
\Box h^2_2= (4G/c^4) B_{0z} \tilde{B}_z.                 %\eqno(13)
\end{eqnarray}
This set may be written in a more symmetric form by defining an energy
density amplitude for both GW and EW [8]. These energy
densities are
\begin{eqnarray}
\rho_{\EW} & = & \displaystyle\frac{\tilde{B}^2_z}{8 \pi}+\frac{\tilde{E}^2_y}
{8 \pi}= \frac{\tilde{B}^2_z}{4 \pi}           \nonumber    \\
\rho_{\GW} & = & \displaystyle\frac{c^2}{16 \pi G} 
[\dot{h}^{2}_{23}+\frac{1}{4} (\dot{h}_{22}-\dot{h}_{33})^2]    %\eqno(14)
\end{eqnarray}
where a dot stands for a time derivative. From the set (14) we may define
each energy amplitude, $\rho_{\EW}\equiv a^2$ and $\rho_{\GW}\equiv b^2$:
\begin{equation}
a \equiv {\tilde{B}_z}/{\sqrt{4\pi}},  \;\;\;\;
b \equiv c \omega h^2_2/\sqrt{16 \pi G}                 %\eqno(15)
\end{equation}
where $h^2_2=-h^3_3=h_{22}$. Substituting these amplitudes to the set of wave
equations, one finds a simpler one:
\begin{equation}
\Box a = (k^2 c^2/\omega^2) \cdot p b \simeq pb,\;\;\;\;
\Box b = pa                                                 %\eqno(16)
\end{equation}
where $p \equiv (2\omega/c^3) B_{0z} (G)^{1/2}$. The last
approximation in the first equation of the above set holds when 
$ k c \simeq \omega$. Therefore, strictly speaking, the photon-graviton 
oscillation is not
exactly symmetric (contrary to what was assumed in Ref.\,[8]).
Moreover, the system should be generalized to take care also of the EW
dispersion law due to the dielectric behaviour of neutral matter or due to
plasma conductivity, as well as quantum electrodynamic corrections to the
Lagrangian (the Feymann box diagrams due to radiative corrections which allow
photon-photon scattering).

\section{Generalized Zel'dovich dispersion law for the          %S3
         graviton-photon oscillation in a refractive medium}

Eqs. (16), describing the photon-graviton oscillations,
can contain different refractive terms; the main ones, due to classical and
quantum electrodynamics, are given by the following wave equation:
\begin{equation}
\Box\vert \tilde{B}_z \equiv \left[ k^2- \left(\mu \epsilon \frac{\omega^2}
       {c^2}+ 4 \pi i \frac{\mu \omega \sigma}{c^2}\right)
      - \left( \frac{e^2}{\hbar c} \right)^2 
       \frac{
        \hbar^3}{m^2c^5} \left( \frac{\omega}{c} \right)^2 B_{0z}^2 \right]
        \tilde{B}_z= B_{0z} k^2 h_2^2                           %\eqno(17)
\end{equation}
where the symbol $\Box\vert$ stands for a generalized D'Alambert operator.

The corresponding EW $\chg$ GW conversion will be no longer
symmetric and the equations must be rewritten as follows:
\begin{eqnarray}
\Box\vert a & \equiv & \Box a -(r_a+r_e+r_B)a= p(k^2c^2/\omega^2)
         b \simeq pb, \nonumber \\
\Box\vert b & \equiv & \Box b - q b = pa                    %\eqno(18)
\end{eqnarray}
where $r_a$, $r_e$ and $r_b$ are the refractive terms related
to those in Eq.\,(17) (which are proportional to $n^2-1$ where $n$ is the
refractive index), connecteed with atomic polarization $(r_a)$, the
plasma conductivity ($r_e$) and the nonlinear behaviour of Q.E.D. ($r_B$).

The similar refractive term $q$ is due to the presence of another form of
energy density perturbed by GWs themselves: $q \simeq (G/c^4) \rho_{em}$.
The refractive terms $r$ and $q$ and the conversion factor $p$ are considered
below in a realistic framework of astrophysical and cosmological interest:
in particular, we analyze the EW $\chg$ GW conversion in
the infrared optical band in laboratory, as well as the reverse phenomenon
[6]. Therefore we shall further discuss:

\begin{description}
\item[(1)] GW emitted by supernovae to the interstellar space and their
            conversion GW $\to$ EW within the kilohertz band;
\item[(2)] EW of the cosmological background radiation spectrum at
         millimeter wavelengths and their deformation due to the conversion
         into GW by cosmological and galactic magnetic fields.
\end{description}

We report here the characteristic values of the refractive terms for
astrophysical and cosmological problems:
{\tiny
\begin{equation}
\begin{array}{lll}
                 &&              \\
Definition & Astrophysical & Cosmological\\
& & \\
r_a \equiv \frac{\omega^2}{c^2}
\left(\frac{\omega^2_{pa}}{\omega_0^2-\omega^2}\right)         &
   \simeq 10^{-40} \left(\frac{n_a}{10^{-1} cm^{-3}}\right)   &
                       =10^{-29} \left(\frac{n_a}{10^{-6} cm^3}\right)\\
      &\;\;\;\; \left(\frac{\omega}{3 \cdot 10^3~\Hz}\right)^2
      \left(\frac{\omega_0}{6 \cdot 10^{14}\Hz}\right)^{-2} cm^{-2} &
       \;\;\;\;    \left(\frac{\omega}{3 \cdot 10^{11}\Hz}\right)^2
        \left(\frac{\omega_0}{6 \cdot 10^{14} \Hz}\right)^{-2} cm^{-2}\\
r_e\equiv -\frac{\omega^2_p}{c^2}=-\frac{4\pi n_e e^2}{m_e c^2}  &
  =-8 \cdot 10^{-14} \left(\frac{n_e}{cm^{-3}}\right)~ cm^{-2}     &
        -8 \cdot 10^{-24} \left(\frac{n_e}{10^{-10} cm^{-3}}\right) cm^{-2}\\
r_{e\pm} \equiv \frac{\omega_p^2}{c^2}
                \frac{\omega}{\omega \pm \omega_B}                         \\
r_{e^+} \equiv \frac{\omega^2}{c^2}\frac{\omega}{\omega+\omega_B}   &
                = 10^{-18} \left(\frac{B}{2G}\right)^{-1} \left(\frac{n_e}
            {cm^{-3}}\right)\left(\frac{\omega}{10^3~\Hz}\right)         \\
r_B\equiv \left(\frac{e}{\hbar c}\right)^2
         \frac{\hbar^3}{m^2c^5}\left(\frac{\omega^2}{c^2}\right)B^2_{z0} &
       =3 \cdot 10^{-44}\left(\frac{B_{0z}}{G}\right)^2
                \left(\frac{\omega}{3 \cdot10^3~\Hz}\right)^2 cm^{-2}     &
     = 3\cdot 10^{-46} \left(\frac{B_{0z}}{10^{-9}}\right)^2
             \left(\frac{\omega}{3 \cdot 10^{11} \Hz}\right)^2 cm^{-2}    \\
q=\frac{G}{c^4}\left(\frac{B^2}{4 \pi} +\rho_r \right)          &
     = 7 \cdot 10^{-51}\left(\frac{B}{\Gs}\right)^2 cm^{-2}   \\
p\equiv 2 \frac{\omega}{c} \frac{\sqrt{G}}{c^2} B_{0z}    &
5 \cdot 10^{-32}  \left(\frac{\omega}{3 \cdot 10^3~\Hz}\right)
                  \left(\frac{B}{\Gs}\right) cm^{-2}                &
=5 \cdot 10^{-33} \left( \frac{\omega}{3 \cdot 10^{11}\Hz}\right)
    \left(\frac{B_0}{10^{-9} \Gs}\right) cm^{-2}  \end{array} \\ %\eqno{(19)}
\end{equation}
} 
where $n_a$ and $n_e$ are, respectively, the neutral gas and electron number
density, $\omega_{\rm pa}=4 \pi n_a e^2/m_a$ is written for
hydrogen, $\omega_0$ is the ionization frequency for hydrogen, $\rho_r$ is
the EM radiation energy density in the propagation medium. All these
quantities are written in the units related to the problem under
consideration.

A generalization of the refractive term $r_e$ ($r_{e \pm}$) is needed
due to the dipolar nature of the propagation of EW in magnetic fields;
the two extreme cases of circular polarized modes (the birefringent modes
[9]) are shown in (19) for the typical
terrestrian magnetic field; $\omega_B=eB/(mc)=1.6 \cdot
10^7 (B/\Gs) \Hz$ and $\omega \sim 10^3
h \ll \omega_B$. These refractive values are a special solution (in the
equatorial plane) of a generalized Appleton-Hartree dispersion equation
[10] whose solutions $r_e (\theta)$ are complicated
functions of the polar angle. 

From a rapid inspection of the conversion term $p$ and the refractive term
$r$ one notices that in general $|r|\gg |p|$,  and the refractive medium
cannot be neglected in principle. From the differential equations (18) and
for the quantities defined in (19) we can evaluate exact and
approximated solutions for the GW $\chg$ EW conversion. We may in
general neglect the GW refractive term $q$ because
\begin{equation}
q \ll p< |r|                                                    %\eqno(20)
\end{equation}
Therefore the set (18) reduces to
\begin{eqnarray}
\Box a & = & r a + k^2 \left(\frac{\omega}{c}\right)^{-2} p b \simeq r a+ pb, 
\nonumber \\
\Box b & = & p a                      %\eqno(21)
\end{eqnarray}
where $r \equiv  r_a+ r_e +r_B$ and $\Box \equiv - \partial^2/\partial
x^2 + c^{-2}\partial^2/\partial t^2$.

The approximation of Eq.\,(21) holds because in general $k \simeq
\omega/c$, i.e., $(\omega/c)^2 \gg r,\ p$. Assuming for the EW and GW
energy density amplitudes a plane wave solution of the form
\begin{equation}
a=a_0 e^{i(kx-\omega t)},\;\;\;\;  b=b_0 e^{i(kx- \omega t)},  %\eqno(22)
\end{equation}
one finds from Eq.\,(21) the exact generalized Zel'dovich dispersion law
\begin{eqnarray}
k^2_{\pm} & = & \frac{\omega^2}{c^2}+\frac{r}{2}
               + \frac{p^2}{2 (\omega/c)^2}  \nonumber \\
 & \pm & \left[\left( \frac{\omega^2}{c^2}+\frac{r}{2}
          +\frac{p^2}{2\omega^2/c^2}\right)^2
- \frac{\omega^2}{c^2}\left(\frac{\omega^2}{c^2}+r\right)\right]^{1/2}.
%\eqno(23)
\end{eqnarray}
This equation has 4 roots: we will restrict ourselves to forward
travelling waves, i.e., consider only positive $k$:
\begin{eqnarray}
k_{\pm} & = & |(k^2_{\pm})^{1/2}| \; = \; \frac{\omega}{c}
\left\{1 + \frac{r}{2\omega^2/c^2}+\frac{p^2}{2\omega^4/c^4} \right. 
\nonumber \\
       & \pm & \left. \left[\left(1 +
\frac{r}{2\omega^2/c^2} + \frac{p^2}{2\omega^4/c^4} \right)^2
-\left(1 + \frac{r}{\omega^2/c^2}\right)\right]^{\frac{1}{2}}
\right\}^{\frac{1}{2}} .
                                                                %\eqno(24)
\end{eqnarray}
Note that negative wave vectors, corresponding to reflected GW, are also
a very interesting phenomenon which may offer a decisive indication of the
EW $\chg$ GW oscillation.
However, the back gravitational reflection, either
in vacuum [4], or in the presence of a refractive medium,
is strongly suppressed as compared with the advancing wave.
Finally, the backward conversion process is no longer quadratically
dependent on the distance $L$ but depends in this way just on the wavelength
$\lambda$ and is consequentely drastically suppressed as compared with
the usual forward conversion. Moreover, the presence of a reflective index
will lead to a ``massive'' photon and to a consequent delay of its arrival
with respect to the corresponding gravitons. A similar delay from the SN
has been considered as a tool for measuring the neutrino mass [11]. Such a
delay would dilute and spoil the EW radio bang signals by supernovae, as
discussed in the Conclusion.

\section{Exact solutions for the generalized Zel'dovich dispersion law}%S4

In general, for advanced waves the energy density amplitudes may be written
as follows:
\begin{eqnarray}
a & = & a_+ e^{i(k_+x- \omega t)}+a_- e^{i(k_- x-\omega t)}, \nonumber \\
b & = & b_+ e^{i(k_+ x-\omega t)}+b_- e^{i(k_- x-\omega t)}.%\eqno(25)
\end{eqnarray}
For $r=0$, $k_{\pm}=
\frac{\omega}{c} \left[1 \pm \frac{p}{(2\omega/c)^2)}\right]$.
Each eigenvalue of the wave vector $k_{\pm}$ in Eq.\,(24)
corresponds to an eigenvector $a_{\pm}$, $b_{\pm}$. To find them,
it is sufficient to consider the approximate Eq.\,(21)
\begin{equation}
\Box a \simeq r a +p b     %\eqno(26)
\end{equation}
valid for $(\omega/c)^2 \gg r,\ p$, i.e., in most real cases. Then
\begin{equation}
a_{\pm}= \frac{p}{\lambda_{\pm}-r}b_{\pm}= \frac{\lambda_{\pm}}{p}b_{\pm}
                                %\eqno(27)
\end{equation}
where $\lambda_{\pm}=k^2_{\pm}-\omega^2/c^2$ are solutions to the
eigenvalue equations (21):
\begin{eqnarray}
\lambda^{ 2}-\lambda~r-p^{ 2} = 0  \nonumber \\
\lambda_{\pm}=\frac{1}{2} r \left(1\pm\sqrt{1+ 4p^2/r^2}\right).  %\eqno(28)
\end{eqnarray}

Let us reconsider two extreme cases: $(\omega/c)^2 \gg p \gg r$
(almost perfect vacuum) and $(\omega/c)^2 \gg r \gg p$
(real refractive medium). The wave vector difference $\Delta k$ is
\begin{eqnarray}
\Delta k & \equiv & k_+- k_- \simeq  \nonumber       \\
 & \simeq & \frac{\omega}{c} \left(\frac{r^2}{4(\omega/c)^4}
   + \frac{p^2}{(\omega/c)^4}+ \frac{p^4}{4(\omega/c)^8}
   +\frac{r p^2}{2(\omega/c)^6}\right)^{1/2}                  %\eqno(29)
\end{eqnarray}
and for $|p| \gg |r|$ one finds the limiting values:
\begin{eqnarray}
k^2_{\pm} & = & \frac{\omega^2}{c^2}\left[ 1+
        \frac{p^2}{2(\omega/c)^4} + \frac{r}{2(\omega/c)^2} \right. 
\nonumber \\
\;\;\;  &  & \left. \pm \frac{p}{(\omega/c)^2}
     \left(1+\frac{r^2}{4 p^2}+\frac{p^2}{4(\omega/c)^4}
        + \frac{r}{2(\omega/c)^2}\right)^{1/2}\right],      
\nonumber \\
\Delta k & \simeq & \frac{p}{(\omega/c)}\left[1+ \frac{r^2}{8 p^2}
        +\frac{p^2}{8(\omega/c)^4}+ \frac{r}{4(\omega/c)^2}\right], 
\nonumber \\
\lambda_{\pm}& = & \pm p\left[1 + \frac{r^2}{4 p^2}\right]^{1/2} +\frac{r}{2}
        =\pm p\left(1+ \frac{r^2}{8 p^2} \right) + \frac{r}{2},  
\nonumber    \\
\frac{\lambda_-}{\lambda_+} & \simeq & \frac{-p+ r/2}{p+r/2}
     \simeq -1 - \frac{r}{p}.                                  %\eqno(30)
\end{eqnarray}
When $|r| \gg p$:,
\begin{eqnarray}
k^2_{\pm} & = & \frac{\omega}{c}\left[1+\frac{r}{2(\omega/c)^2}+
 \frac{p^2}{2(\omega/c)^4} \right.     
\nonumber                                \\
\;\;\; & & \left. +\frac{p^2}{2r(\omega/c)^2} + 
\frac{p^2}{8(\omega/c)^2}\right],      \nonumber \\
\Delta k & \simeq & \frac{|r|}{2(\omega/c)}
        \left[1+ \frac{2 p^2}{r^2}+\frac{p^4}{2(\omega/c)^4 r^2}
        + \frac{p^2}{r(\omega/c)^2}\right],                        %\eqno(31)
\end{eqnarray}
then for $r > 0$
\begin{eqnarray}
\lambda_{\pm}= \frac{r}{2} \left[1 \pm \left(1+\frac{4p^2}{r^2}\right)^{1/2}
     \right] \simeq \left\{
\begin{array}{ll} r\left(1+\frac{p^2}{r^2}\right),&  \\
             -\frac{p^2}{r},&                
\end{array} 
\right.
               %\eqno(32)
\end{eqnarray}
while for $r < 0$
\begin{equation}
\lambda_{\pm}=\frac{|r|}{2}
                   \left[-1 \pm \left(1+\frac{4p^2}{r^2}\right)^{1/2}\right]
     \simeq \left\{
\begin{array}{l} \frac{p^2}{|r|},        \\
                  -|r|- \frac{p^2}{|r|};  
\end{array}
\right.                    %\eqno(33)
\end{equation}
\begin{eqnarray}
\frac{\lambda_-}{\lambda_+} & \simeq &
\frac{-p^2/r^2}{1 + rp^2/r^2} = \frac{-p^2}{p+r^2}
                                 \;\;\;\; for \;\; r>0, \nonumber \\
\frac{\lambda_-}{\lambda_+} & \simeq &
    \frac{-|r|(1+p^2/r^2)}{p^2/|r|} =     \nonumber    \\ 
          & & = \frac{-r^2}{p^2} \left( 1+\frac{p^2}{r^2} \right)=
         -\frac{r^2}{p^2}-1,            \; for \;\;  r<0.%\eqno(34)
\end{eqnarray}
Given these expressions for $\Delta k$, $\lambda_{\pm}$, and the eigenvector
relations of Eq.\,(27) in the needed approximations, we can easely analyze
the GW $\chg$ EW conversion for any realistic framework in
physics, astrophysics and cosmology.

\section{Conversion efficiency in a single oscillatory period}   %S5
Let us first consider as initial conditions ($t=0$, $x=0$) a vanishing GW
($b_{in}=0$) and a strong EW beam propagating in an orthogonal stationary
field $B_{0z}$ ($a_{in} \neq 0$) [6]; for instance, the EW may be a laser
beam along a magnetized tunnel where $B_{0z} \parallel \tilde{B_z}$, i.e.,
 where the stationary field is parallel to the polarized EW $\tilde{B}$
field. The corresponding GW energy density amplitude $b$ is zero at $t=0$,
$x=0$, while the EW amplitude $a$ has a given value $a_{in}$. From Eq.\,(25)
one obtains:
\begin{equation}
b_+=- b_-,\quad a_{\pm}= \frac{\lambda_{\pm}}{p} b_{\pm},\quad
\frac{a_+}{a_-}=- \frac{\lambda_+}{\lambda_-}.       %\eqno(35)
\end{equation}
From Eq.\,(25) we thus obtain
\begin{eqnarray}
a & = & a_+\left[\eip  - \frac{\lambda_-}{\lambda_+} \eim\right] , \nonumber \\
b & = & b_+\left[\eip - \eim \right] = \nonumber \\
  & = & 2a _+\frac{p}{\lambda_+}
     i\left[\sin \frac{\Delta\kappa x}{2}\right]\ei          %\eqno(36)
\end{eqnarray}
where $k_{\pm}$, $\lambda_{\pm}$, $\Delta k$ are defined in (24),\ (28),
\ (29); in the case $p\gg |r|$ we obtain from Eq. (30) the explicit solutions
\begin{eqnarray}
a & = & 2 a_+ \cos \frac{\Delta k x }{2} \ei + \frac{r}{p}\,a_+ \eim , 
\nonumber \\
b & = & \frac{2a_+ i}{1+r/(2p)}\sin\frac{\Delta k x}{2} \ei .     %\eqno(37)
\end{eqnarray}
For $r=0$ the solution of Eq.\,(37) reduces to the perfect vacuum case:
\begin{equation}
a=2 a_+ \cos \frac{\Delta k x }{2} \ei, \;\;\;
b= 2a_+i\, \sin \frac{ \Delta k x}{2}\ei .           %\eqno(38)
\end{equation}
In the low refraction limit ($p \gg |r|$) the EW $\to$ GW conversion
occurs at a rate
\begin{equation}
\alpha \equiv \frac{|b|^2}{|a|^2}
= \frac{4}{(1+r/2p)^2} \frac{\sin^2 (\frac{1}{2}\Delta K\,x)}
{\left[4 \cos^2 (\frac{1}{2}\Delta k\,x)
 +4 (r/p) \cos(\frac{1}{2}\Delta k\,x) + r^2/p^2\right]};         %\eqno(39)
\end{equation}
in the zero refraction limit
\begin{equation}
\lim_{r \to 0} \alpha = \tan^2{\left(\frac{\Delta k~x}{2}\right)} \simeq
\frac{p^2 x^2}{4(\omega/c)^2}= \frac{G}{c^4} B^2_{0z} x^2      %\eqno(40)
\end{equation}
where the last approximation in Eq.\,(40) holds for $x \leq  \Lc=
{ 2 \pi/\Delta k}=(2 \pi/p) (\omega/c)$. Therefore
we return to the classical conversion efficiency of Eqs. (1),(2). 

From Eq.\,(39) one may suspect that the conversion efficiency for $r \gg p$
will be suppressed by a factor $(p/r)^2$, but this is not the
case because Eq.\,(39) holds only under the assumption $p\gg |r|$. Indeed,
for $|r| \gg p$ we should consider the two extreme possibilities $r < 0$ or
$r > 0$ (i.e., $|r_e| < |r_a+r_B|$ \ or \ $|r_e| > |r_a+r_B|$). 
From the approximate eigenvector and eigenvalue expressions
$\lambda_{\pm}$, $\Delta k$, $\lambda_-/\lambda_+$ in
Eqs.\,(31)--(36) we easily find the EW and GW energy amplitudes, $a$ and
$b$, respectively: for $r > 0$ and $p \ll r$
\begin{eqnarray}
a & = & a_+ \left[ \eip +
     \frac{p^2}{p^2+r^2}\eim \right], \nonumber    \\
b & = & 2i \frac{pa_+}{\lambda_+}\sin \frac{\Delta k x}{2} \ei
  \nonumber \\
  & = & 2 \frac{p}{r} \frac{a_+i}{(1+p^2/r^2)}
   \sin\frac{\Delta k x}{2}\ei ,                   %\eqno(41)
\end{eqnarray}
while for $r < 0$ and $p \ll |r|$
\begin{eqnarray}
a & = & a_+\Bigl[\eip  + \frac{p^2+ r^2}{p^2}\eim \Bigr] = \nonumber \\
  & = & \frac{r^2}{p^2}a_+
  \Bigl[ \frac{p^2}{r^2} \eip + \frac{p^2+ r^2}{p^2} \eim \Bigr]; \nonumber \\
b & = & 2i \frac{pa_+}{\lambda_+}\sin \frac{\Delta k x}{2} \ei = \nonumber \\
  & = & 2 \frac{|r|}{p} a_+ i\,\sin\frac{\Delta k x}{2}\ei    .%\eqno(42)
\end{eqnarray}
The energy conversion ratio for small distances
$\; x \ll \frac{2\pi}{\Delta k}=\Lc
\simeq \frac{4 \pi}{|r|}\omega/c$ is given by two $\alpha$ values:
\begin{eqnarray}
\mbox{(i) for $r >0$:} & 
\alpha= \frac{|b|^2}{|a|^2}=
\frac{4 \frac{p^2}{r^2} \sin^2\frac{\Delta k x}{2} }
{\Big(1+ \frac{p^2}{r^2}\Big)^2[1+\Big(\frac{p^2}{p^2+r^2}
\Big)^2+\frac{2 p^2}{p^2+r^2} \cos{\Delta k x}]};\\            %\eqno(43)
\mbox{(ii) for $r< 0$:} & 
\alpha= \frac{|b|^2}{|a|^2}=\frac{\frac{4p^2}{r^2}\sin^2{\Big(
\frac{\Delta kx}{2}\Big)}}{[
\frac{p^4}{r^4}+\Big(\frac{p^{ 2}+r^2}{r^2}\Big)^2+2
\frac{p^2}{r^2}\Big(\frac{p^2+r^2}{r^2}\Big) \cos{\Delta kx}]} %\eqno(44)
\end{eqnarray}
Apparently the ratio is suppressed by a factor $p^2/r^2$.

However, in both solutions the conversion ratio may be approximated for
$p/r\to 0$ as follows:
\begin{equation}
\alpha= \frac{|b|^2}{|a|^2}\simeq 4 \frac{p^2}{r^2}
  \sin^2 \frac{\Delta kx}{2} \simeq 4\frac{p^2}{r^2}
  \ \frac{r^2}{16 (\omega/c)^2} =\frac{G}{c^4}B^2_{0z}x^2,   %\eqno(45)
\end{equation}
i.e., even when $|r| > p$, but as long as $x \ll 2\pi/\Delta k$, the
conversion factor is as large as in the perfect vacuum case ($|p| \gg |r|$).
This surprising result may be understood as follows: in vacuum the
conversion ratio is not suppressed but the corresponding
coherence length $2 \pi/\Delta k$ (for full conversion) is very large:
\begin{equation}
\Lc(p\gg r)= \frac{2\pi}{\Delta k}= \frac{2\pi}{p}\,\frac{\omega}{c}
=\frac{\pi c^2}{\sqrt{G}B_{0z}}
   =1.3 \cdot 10^{19} \Bigl(\frac{B_{0z}}{10^6 G}\Bigr)^{-1}\ cm %\eqno(46)
\end{equation}
In the presence of a refractive term $r$ the conversion efficiency $\alpha$
is suppressed by a large factor $(p/r)^2$ but the coherence
length (for full conversion) is now much shorter as compared with the
previous one just by the same correcting factor:
{\small
\begin{equation}
\Lc(p \ll |r_{ e}|)= \frac{2 \pi}{\Delta k}
 =\frac{4\pi}{r}\,\frac{\omega}{c} =2 \frac{p}{r} \Lc(r{=}0)
 = 4 \cdot 10^{4}\biggl(\frac{\omega}{3 \cdot 10^{4}\Hz}\biggr)^{-1}
                 \biggl(\frac{B_{0z}}{G}\biggr)^{-2} \ {\rm km}. %\eqno(47)
\end{equation}
}
This effect exactly compensates the suppression factor $(p/r)^2$
in the efficiency, leading to the same result as is valid in vacuum
(Eqs.\,(39), (40)). In summary, the EW $\chg$ GW conversion for
$x \leq \Lc(r)$ ignores the refraction, even if $r \gg p$.
However, for $x \gg \Lc$, we should expect a new phenomenon due to
wave packet separation: multiple EW $\chg$ GW conversion.

\section{Multiple EW $\chg$ GW conversion}    %S6

The amplitude solutions of the previous section are no longer valid at
distances $x \gg \Lc$. Indeed, as long as $|a| \ll |b|$
(or {\it vice versa\/}) the conversion is a monotone (rather
than oscillatory) phenomenon which we call ``multiple conversion''. Let us
describe this ``multiple conversion'' using an analogy: a rich man meets a
very poor friend during his walk. The two friends agree to play, while
walking, a very innocent game, ``the give and take game'': at each finite
distance ($\Lc$), each of them gives the other a small fraction, say, $4
p^2/r^2$, of his own pocket money at that moment. Let us label $|a|^2_0$
the total initial venture capital (pocket money) of the rich man and
$|b|^2_0=0$ the corresponding initial ``no money'' of the poor friend.
After the first distance $\Lc$ (see Eq.\,(47)) the rich man gives
$4p^2/r^2 |a|^2_0$ to his friend, so that the rich man remains
with $(1-4 p^2/r^2)|a|^2_0$ (the poorer one gains at the first
step $4 p^2/r^2 |a|^2_0$). At the next step $\Lc$ the rich one
offers a similar fraction of money
$4 (p^2/r^2)(1 -4p^2/r^2)|a|^2_0$ but receives back only $
(4p^2/r^2)^2 |a|^2_0$. On the contrary, the poor one will capitalize
$8(p^2/r^2)(1-2 p^2/r^2)$. As long as
$p/r\ll 1 $, ``the oscillatory give and take game'' works one-
way as a monotonically irreversible process (the poorer one will become
richer and the rich will become poorer). Only when both players hold an
equal capital, the process will go on symmetrically, in a reversible
oscillatory way. To be more quantitative, let us consider the energy
density evolution of the EW and GW. From Eq.\,(36) we can redefine the
amplitude $a_+$ as follows:
\begin{equation}
a_+ = \left\{
\begin{array}{ll}
\frac{p^2+r^2}{2p^2+r^2}~a_{\rm in}, & r>0; \\
a_+=\frac{p^2}{2p^2+r^2}~a_{\rm in}      & r<0.   
\end{array}  
\right.   %\eqno(48)
\end{equation}
The corresponding energy densities $\rho_{\EW}$ and $\rho_{\GW}$ become
\begin{eqnarray}
\rho_{\rm EW}=|a|^2=a_{\rm in}^2 \Big(1 - \frac{2 p^2}{2 p^2+ r^2}\Big)^2
\biggl[1+ \Big(\frac{p^2}{p^2+r^2}\Big)^2
  +\frac{2 p^2}{p^2+r^2} \cos{(\Delta k x)}\biggr] \;\; r > 0, 
\nonumber \\
\rho_{\rm EW}=|a|^2=a_{\rm in}^2 \Big( \frac{ p^2}{2 p^2+ r^2}\Big)^2
   \biggl[1+ \Big(\frac{p^2}{p^2+r^2}\Big)^2+\frac{2(p^2+r^2)}{p^2}
      \cos{(\Delta k x)}\biggr], \;\;  r<0.           %\eqno(49)
\end{eqnarray}
When $\Delta k~x= \pi$, i.e., for $x=\Lc/2=\pi/\Delta k$, one gets
\begin{eqnarray}
\rho_{\rm EW} & = & \rho_{\rm in}\Big(1 - \frac{2 p^2}{2 p^2+ r^2}\Big)^2
\Big(1- \frac{p^2}{p^2+r^2}\Big)^2\simeq
  \Big(1- \frac{4 p^2}{r^2}\Big) \rho_{\rm EW\,in}, \;\; r >0; \nonumber \\
\rho_{\rm EW} & = & \displaystyle \rho_{in}\Big(\frac{2 p^2}{2 p^2 +r^2}\Big)^2
\Big(1-\frac{p^2+r^2}{p^2}\Big)^2  \simeq
\Big(1- \frac{4 p^2}{r^2}\Big) \rho_{\rm EW\,in}, \;\; r<0. %\eqno(50)
\end{eqnarray}
In a similar way one easily finds from Eqs.\,(43), (44), either for $r>0$ or
$r< 0$ after a distance $x= \Lc/2$:
\begin{equation}
\rho_{\GW}=|b|^2 \simeq 4 \frac{p^2}{r^2} \rho_{\rm EW\,in}. %\eqno(51)
\end{equation}
Therefore, adding Eqs.\,(50) and (51), we easily verify the energy
conservation law $\rho_{\rm in}=\rho_{\rm EW}+\rho_{\rm GW}$. If we want to
generalize the process, we may define an average energy density
$A_n=|a_n|^2$, $B_n=|b_n|^2$, i.e. the energy densities at any step $n$; the
energy density evolution will have the form (from generalized
Eqs.\,(50), (51)):
\begin{equation}
A_n - A_{n-1} =- \frac{4 p^2}{r^2}\,A_n +\frac{4 p^2}{r^2}\,B_n,
\hspace{15mm}
B_n - B_{n-1}=  \frac{4 p^2}{r^2}\,A_n -\frac{4p^2}{r^2}\,B_n.%\eqno(52)
\end{equation}
Dividing by $\Lc/2$ and assuming a continuous limit, we can
write the following set of equations:
\begin{eqnarray}
  \frac{d A}{dx} & \equiv & \frac{A_n {-} A_{n-1}}{\Lc/2}=
     - 8 \frac{p^2}{r^2 \Lc} (A{-}B)=-\frac{1}{L}(A{-}B), \nonumber \\
\frac{d B}{dx} & \equiv & \frac{B_n {-} B_{n-1}}{\Lc/2}=
     8 \frac{p^2}{r^2 \Lc} (A{-}B)=\frac{1}{L}(A{-}B)  %\eqno(53)
\end{eqnarray}
where the characteristic distance $L\equiv (r^2/8p^2)\Lc$ is the
relaxation length of the system (53). Assuming a solution of the form
$A=A_0 e^{kx}$, $B=B_0 e^{kx}$, one finds the eigenvectors: $k_+ =0$,
$k_-=-2/L$, not to be confused with the previous ones. For the initial
boundary conditions $A(t=0, x=0)=A_0$, $B(t=0, x=0)=0$ one finally obtains:
% \twocolumn \bls{0.96} \noi
\begin{eqnarray}
A & \equiv & |a|^2 =\frac{1}{2} A_0(1+ e^{-2x/L}) \simeq A_0 (1- x/L), 
\nonumber \\
B & \equiv & |b|^2 =\frac{1}{2} A_0(1- e^{-2x/L}) \simeq A_0 \frac{x}{L},  
\nonumber \\
\alpha & = & B/A= \tanh(x/L) \simeq x/L \simeq
          8 (p^2/r^2) (x/\Lc).                  %\eqno(54)
\end{eqnarray}
The last approximations hold as long as $(r^2/p^2) \Lc>x > \Lc$.
The average energy densities $A$ and $B$ are modulated by the $|a|$ and
$|b|$ amplitudes in Eqs.\,(36)--(42) but their average value evolution
in any period is given by (54). When $A \simeq A_0/2$, $B\simeq B_0/2$, the
oscillatory behaviour occurs in a reversible form.  The multiconversion has
an analogous phenomenon in cosmology when left-handed neutrinos with Dirac
and Majorana masses oscillate only in one way, leading to thermalization
of the right-handed sterile neutrinos in the early Universe [12].

\section{Oscillatory coherence lengths}                       %S6

By (46) and (47), the characteristic coherence lengths in astrophysical and
cosmological problems become
\begin{eqnarray}
\Lc(r \ll p) & = & \frac{2 \pi}{p}\frac{\omega}{c}=4 \cdot 10^8 \Big(
        \frac{B}{10^6 G}\Big)^{-1} {\rm sec}~c, \nonumber \\
\Lc(p\ll r_{B}) & = & \frac{4 \pi}{|r|}\frac{\omega}{c} 
= 4 \cdot 10^{14} \Big(\frac{\omega}{3 \cdot 10^{14} \hbar c}\Big)^{-1}
                \Big(\frac{B}{10^6 G}\Big)^{-2}, cm  \nonumber \\
\Lc(p \ll |r_{ e}|) & = & 1.5 \cdot 10^2 \frac{\omega}
 {3 \cdot 10^3 \Hz} \Big(\frac{n_e}{cm^{-3}}\Big)^{-1}~{\rm km} \nonumber \\
   & & = 5\cdot 10^{ 4} \frac{\omega}{3\cdot 10^{11}\Hz}
       \frac{n_{ e}}{cm^{ -3}}\ c\cdot {\rm sec},  \nonumber   \\
\Lc(p \ll |r_{e+}|) & = & 1.2 \cdot 10^7 \frac{\omega}{3 \cdot 10^3\Hz}
     \Big(\frac{n_e}{cm^{-3}}\Big)^{-1}\frac{B}{2~\Gs} {\rm km}. %\eqno(55)
\end{eqnarray}
The coherence length is defined by the minimum distance between the coherence
lengths and the inhomogeneity scales of the stationary fields, magnetic or
electric.
For instance, we may also consider the multiconversion efficiency due to the
atomic electric fields inside normal matter, where the photon and graviton
wavelengths are much smaller than the atomic radius: $\lambda_{\gamma}$,
$\lambda_{\tilde{g}} \ll \AA$.
The random nuclear electrostatic fields become external stationary fields
responsible for the EW $\chg$ GW conversion of high energetic
photons (and gravitons). The coherence length is just the atomic radius at
normal densities. As a first approximation, in hydrogen, the eletric field
is on average $3 \cdot 10^7 {\rm V/cm}$, corresponding to $10^5$ Gauss. For a
coherence distance $\Lc \simeq 1 \AA \simeq 10^{-8}$ cm and for a
conversion distance $x$ of $10^7$ cm one obtains:
\begin{equation}
\alpha \simeq 10^{-41} \Big(\frac{x}{\rm 100~km}\Big) Z^2    %\eqno(56)
\end{equation}
%\twocolumn \noi
where Z is the matter atomic number. Therefore in normal matter the GW
$\chg$  EW conversion is much weaker than the coherent one
for any laboratory field (in Eq.\,(1)) and can be neglected. For higher
densities, as those in neutron stars just before neutronization, one
finds much smaller coherence lengths but higher electric fields between
nucleons: $\Lc \simeq 10^{-13}$ cm, $ E \simeq 3 \cdot 10^{17} {\rm
V/cm}$. Therefore
\begin{equation}
\alpha \simeq 10^{-26}\ \frac{x}{\rm 100\ km}. %\eqno(57)
\end{equation}
High energetic photons ($E_{\tilde{\gamma}} \geq 100$ Mev) will generate a
small fraction of high energetic gravitons. We may consider incoherent
multiconversion of thermal photons in thermal equilibrium in a bath of
electron (or fermion, or boson) pairs (in a hot stellar core). During a
supernovae explosive stage the core temperature may reach a value
$kT_{\gamma}\geq 10-100$ MeV. This occurs in a region where neutrinos reach
opacity (the so-called neutrino photosphere). Electron and neutrino pairs for
a fraction of a second are in thermal equilibrium $T_{\gamma} \simeq T_{\nu_e,
\tilde{\nu_e}} \simeq T_{e_{\pm}}$. The electron pairs (as well as photons
and neutrinos) are trapped in the stellar core and are forced to run for a
fraction of a second in a finite random walk for the total distance, let
us say, of $x \simeq ct \simeq 3 \cdot 10^{10}$ cm. Therefore the
multiconversion efficiency for thermal photons (into gravitons) in the
presence of the electric field $E$ of dense electron pair fields, at a
minimal distance $r_{\rm min} \simeq h/kT_{\gamma} \simeq \lambda_{\gamma}
t$, is
\begin{equation}
<\alpha> \simeq \frac{G}{c^4} <E>^2 \Lc x     
\simeq 2.4 \cdot 10^{-19} \Big(
\frac{T}{\rm 10\ MeV}\Big)^3 \; \frac{x}{c~{\rm sec}}.       %\eqno(58)
\end{equation}
We learn herefrom that during the SN1987A explosion
a small fraction of internal energy of neutrino pairs (and photons),
$E_{\nu} \geq 10^{53}$ erg, is converted into a source of highly energetic
gravitons $<E_{\tilde{g}}> \simeq 10-100$ MeV with the corresponding
total energy burst
\begin{equation}
E_{\tilde{g}}\simeq <\alpha> E_{\nu\ {\rm SN}}
        \simeq 2.4\cdot 10^{34}\ \ {\rm erg},      %\eqno(59)
\end{equation}
i.e., a power in gravitons comparable to our solar power in EM waves
(per second) but not so easily detectable (and much below the power of gamma
burst sources). It is suggestive to consider this phenomenon as a possible
explanation for energy transfer outside the collapse when neutrino opacity
occurs. This result may lead to a new approch to SN explosion
models.  In cosmology the multiconversion may also occur at high densities
and energies in the early Universe. However, the time scale of the
phenomenon and therefore the corresponding flight time are also related
to the temperature, namely, that in a radiation dominated Universe:
$t \simeq 1$ sec $(T/{\rm MeV})^{-2}$. From Eq.\,(58) one gets:
\begin{equation}
\alpha \simeq 2.4 \cdot 10^{-22} \Big(\frac{T}{MeV}\Big)^3\ \frac{t}{\rm sec}
      =2.4 \cdot 10^{-22}\ \frac{T}{MeV}.      %\eqno(60)
\end{equation}
In a radiation dominated Universe complete conversion occurs at a critical
temperature
\begin{equation}
T \geq 10^{22}\ {\rm MeV}= 10^{19}\ {\rm GeV} = m_{\rm pl} c^2. %\eqno(61)
\end{equation}
As we should expect from dimensional arguments, the photon-graviton
conversion is a totally efficient process only as early as at Planck times.
Therefore the photon-graviton conversion may be at least a key process in
keeping gravitons in thermal equilibrium in the Universe at very early epochs.

\section{Cosmic background multiconversion by cosmological
            and galactic magnetic fields}                            % S7
A scenario with the conversion process (EW $\to$ GW) playing a relevant
physical role has been pointed out by Zel'dovich [8]
in cosmology. He considered coherent conversion of the background
radiation (CBR) into GW (the opposite conversion is clearly negligible) by a
cosmological magnetic field $B_0=10^{-9} \div 10^{-6}$ Gauss. The author
[8] studied a refractive medium but took into account only single
conversion (where the growth is quadratic or square sinusoidal with the
distance and has an upper limit $\simeq p^2/r^2$; see
Eq.\,(40)). In particular, at the redshift $z = 10^3$, for a cosmological
magnetic field $B_0=1$ Gauss, in the presence of a baryon density
$n_a=10^3~cm^{-3}$, a plasma density $n_e=10^{-1} cm^{-3}$ and for a
characteristic EW wave vector $k=\omega/c=10^4~cm^{-1}$, Zel'dovich found
\begin{equation}
\alpha_{\max} \simeq 4 p^2/r^2= 4\cdot 10^{-12}.    %\eqno(62)
\end{equation}
He concluded that $\alpha_{\max}$ in Eq.\,(62) is an absolute upper limit
for GW $\chg$ EW conversion. However, this result holds only for one
oscillation over a coherence length $\Lc$:
\begin{equation}
\Lc(p < |r_e|)=\frac{4 \pi}{|r|}\frac{\omega}{c} = 
    16~\frac{\omega}{3 \cdot 10^{14} \Hz}
    \Bigl(\frac{n_e}{10^{-1}cm^{-3}}\Bigr)^{-1}~{\rm l.y.}  %\eqno(63)
\end{equation}
The cosmological age at recombination near the redshift $z=1000$ is
$t=t_0(1+z)^{-3/2} \simeq 5 \cdot 10^5$ yrs. Therefore the cumulative
multiconversion (see Section 7) takes place nealy $3 \cdot 10^4$ times
and leads to a total conversion factor
\begin{equation}
\alpha \simeq \cdot 10^{-7}.                                  %\eqno(64)
\end{equation}
This value is still small but almost at a detectable level. However, it may
seem exaggerated (and unrealistic) to consider such a present primordial
coherent cosmological magnetic field $B_0=10^{-6}$ Gauss. It is,
on the contrary, quite realistic to consider an incoherent magnetic field
at smaller (galactic) scales, actually the observed random galactic field,
at values of $10^{-6} \div 10^{-5}$ Gauss. At redshift $z=10^3$,
$\Lc \simeq 10$ l.y., a value which is just comparable with the
observed homogeneous scale in the Galaxy
and the coherence lengths derived for the refractive index in Eq.\,(63).
Therefore the conversion factor in Eq.\,(64) is realistic,
related to the inhomogeneous random galactic field of the interstellar space
at recombination.
The primordial galactic contrast over angular scales $\theta \simeq 3^{"}$,
needed for adiabatic galaxy formation, is in CDM models of the order
\begin{equation}
\frac{\Delta T}{T} \Bigr|_{z=10^3} \geq 10^{-6}.        %\eqno(65)
\end{equation}
However, this inhomogeneity has not yet been observed; one possibility is that
random multiconversion of the 3 K CBR into GBR leads to a smoother random
temperature contrast:
\begin{equation}
\frac{\Delta T}{T}\Bigr|_{z=10^3} = N \frac{p^2}{r^2}
    \pm \sqrt{N}\frac{p^2}{r^2}\simeq 10^{-7}\pm 10^{-9}    %\eqno(66)
\end{equation}
where we assume an average present-day galactic magnetic field to be
$3 \cdot 10^{-6}$ Gauss.
Moreover, the conversion of photons into gravitons may deplete the original
spectrum leading to a smaller effective CBR at higher frequency. A relevant
consequence of photon-graviton multiconversion is therefore a possible
presence of a comptonization factor $y$ at a level of few $10^{-7}$, much
smaller (yet) than the present bounds by COBE ($y \sim 10^{-5}$).

\section{Conclusions}

As mentioned in the Introduction, the best astrophysical opportunity to test
the GW $\chg$ EW conversion lies in the huge GW release by supernovae at
kilohertz (or tens of kilohertz) frequencies when these waves cross the
nearest local magnetic field. Let us assume a nominal SN explosion at 50 kpc
whose total GW energy is $10^{51}$ erg ($1\%$ of the
neutrino burst). The corresponding flux energy is
$\Phi= 4\cdot 10^3$ erg cm$^{-2}$. Under this assumption
the total GW $\to$ EW conversion may occur either
by the terrestrial, Jovian, solar,
interstellar, intergalactic magnetic fields, or in SN or its surroundings,
either coherently or, more often, incoherently,
following the discussion of Eq.\,(55):
\begin{eqnarray}
\alpha_{\oplus +} & \simeq & 2\cdot 10^{-32}\Big(\frac{B}{0.5 \Gs}\Big)^2
               \Big(\frac{L}{10^4 {\rm km}}\Big)^2,  \nonumber   \\
\alpha_{J} & \simeq & 1.3\cdot 10^{-28}\Big(\frac{B}{4\Gs}\Big)^2
              \Big(\frac{L}{10^5 {\rm km}}\Big)^2,  \nonumber    \\
\alpha_{\odot} & \simeq & 1.3 \cdot 10^{-26}\Big(\frac{B}{4\Gs}\Big)^2
              \Big(\frac{L}{10^6 {\rm km}}\Big)^2,  \nonumber \\
\alpha_{\rm i.g.inch} & \simeq & 8\cdot 10^{-16}
              \Big(\frac{B}{10^{-6}\Gs}\Big)^2
  \frac{L}{100 {\rm kpc}}\ \frac{L_c}{100 {\rm pc}}, \nonumber \\
\alpha_{\rm i.g.coh} & \simeq & 
8 \cdot 10^{-13}\Big(\frac{B}{10^{-5}\Gs}\Big)^2
              \Big(\frac{L}{100 {\rm kpc}}\Big)^2, \nonumber   \\
\alpha_{NS} & \simeq & 8 \cdot 10^{-12} \Big(\frac{B}{10^{12}\Gs}\Big)^2
              \Big(\frac{L}{10^2 {\rm km}}\Big)^2.          %\eqno(67)
\end{eqnarray}
The coherence size of galactic magnetic fields has been discussed above (see
Eqs. (54),(55)). However, the last and largest conversion
for the SN occurs in an enriched plasma arund the SN, so  the EWs will
be easily screened at those kilohertz radio frequencies.
The possible energy enhancement by inverse Compton
scattering of GeV electrons by SN and the possible emission
of consequent shorter waves could be able to overcome the
refraction and reflection in the
ionized domains, but these signals will not be discussed here.
Therefore we shall neglect the role of the kilohertz GWs converted near a SN.
For the same reason we may neglect the GW $\to$ EW conversion near our Sun.
We may consider as first conversion to be near (outside) the Earth, i.e.,
the earliest to be observable, in this case the total energy flux will be
\begin{equation}
 \Phi_{\rm EW}=\int \frac{d\Phi}{d\omega} d\omega
    =8\cdot 10^{-29}\Big(\frac{B}{B_{\oplus}}\Big)^2
    \Big(\frac{L}{10^4 {\rm km}}\Big)^2~~ \frac{\rm erg}{cm^2}.
                                              %\eqno(68)
\end{equation}
The total spread of the GW frequency, being probably characterized by a
flat step spectrum, falls within $10^4 \div 10^5$ \Hz\ band
and the consequent flux is:
\begin{equation}
 \frac{d\Phi_{\rm EW}}{d\omega}\biggr|_{SN 1987A}
=8\cdot 10^{-33}\frac{\rm erg}{cm^2\,\Hz} = 8\cdot 10^{-4}~\mu\Jan,
                              \\                   %\eqno(69)
\end{equation}
too low to be observable. For a 10 kpc source the differential flux would be
much larger:
\begin{equation}
\frac{d\Phi_{\oplus~{\rm EW}}}{d\omega}\Bigr|_{\rm 10 kpc}
           =2\cdot 10^{-2}~ \mu \Jan.                          %\eqno(70)
\end{equation}
For the same source a better observational place is near Jupiter's orbit where
the total flux would be nearly 6400 times greater, leading to a radio burst:
\begin{equation}
\frac{d\Phi_{\rm J~EW}}{d\omega}
        =1.38\cdot 10^2~\mu\Jan=0.14~{\rm m}\Jan,                 %\eqno(71)
\end{equation}
within the present radio sensitivity ranges.

Finally we shall reject the overoptimistic case of a total coherent
intergalactic field and restrict ourselves to the more modest (but
more realistic) case of an incoherent field at 100 pc coherence length.
In that case, nevertheless, for a source like SN 1987A, the total energy flux
will be still impressive:
\begin{equation}
\frac{d\Phi_{\rm EW}}{d\omega}=3.2\cdot 10^{-16}
     ~\frac{\rm erg}{cm^2\,\Hz}=3.2\cdot 10^7~ \Jan.          %\eqno(72)
\end{equation}
Unfortunately, the flux will not be a longer ``prompt'' one (as is near
Jupiter) but is very much delayed in time because those low energy
``photons'' behave like massive particles with a relativistic Lorentz factor
\begin{equation}
\Gamma_\gamma=\frac{E_\gamma}{E_{\rm pl}}\simeq
  \frac{\omega}{\omega_p}
  \simeq 10~\frac{\omega}{10^5 \Hz}\ \frac{n_e}{1~cm^{-3}}.  %\eqno(73)
\end{equation}
Consequentely, the ``fastest'' energetic EWs (possibly at 30--100 kHz for
neutron-star or black-hole sizes) will reach the Earth with a large time
delay $\tau_d$ with respect to the prompt SN events:
\begin{equation}
\tau_d=\frac{1}{2\Gamma_\gamma^2}\frac{L}{c}\simeq 750~
       \frac{L}{1.5\cdot 10^5 {\rm yrs}}
          \Big(\frac{\omega}{10^5 \Hz}\Big)^{-2}~{\rm yrs}.   %\eqno(74)
\end{equation}
This delay would introduce a huge time dilution of the signal, as well as a
severe flux reduction
\begin{equation}
\frac{d\Phi_{\rm EW}}{d\omega\ dt}\simeq 10^{-3}\ \Jan      %\eqno(75)
\end{equation}
in the above estimates.
Moreover, a large ``refractive'' index will give life to some random walk of
the radio signals and will smear out their directionality more and more,
leading to a further spread and dilution of the signal/noise ratio.
Neverthless, the tens-of-kilohertz radio wave band is a very exciting
``astrophysical'' band to be considered for discovering the secret of
gravitational waves. Unfortunately, noise is large and little is known to the
author at these bands. These frequencies are actually already used to
discover other kinds of secrets. The same satellites which probably look
into the deep blue sea at those frequencies might have already recorded,
without being aware of it, both the prompt and the delayed signals due to SN
1987A GW $\to$ EW conversion (if their sensitivity is above the noise).
Finally, it sounds ironically that these military satellites could hide in
their recorded data at the lowest radio energies just the exciting secrets
of the most powerful explosions in our Galaxy.

\section*{Acknowledgement}
I wish to thank Prof. A. Dolgov and Dr. R. Bolzanello for support and
enlightening discussions and Dr. A. Salis for deep reading and
comments on the article.

\end{document}